\documentclass[a4paper,11pt]{article}
\pdfoutput=1 
\usepackage{jheppub} 
\usepackage[T1]{fontenc} 
                     
\usepackage[font=small]{caption}
\usepackage{subcaption}
\usepackage[title]{appendix}

\newcommand{\be}{\begin{equation}} 
\newcommand{\ee}{\end{equation}} 
\newcommand{\bea}{\begin{eqnarray}}  
\newcommand{\eea}{\end{eqnarray}}
\newcommand{\bs}{\begin{split}} 
\newcommand{\es}{\end{split}}
\newcommand{\nn}{\nonumber \\ } 

\title{\boldmath Surveying the Scope of the $SU(2)_L$ Scalar Septet Sector}

\author{C. Alvarado}
\author{L. Lehman}
\author{and B. Ostdiek}

\affiliation{Department of Physics, University of Notre Dame, \\225 Nieuwland Science Hall\\ Notre Dame, IN 46556\\ USA}

\emailAdd{calvara1@nd.edu}
\emailAdd{llehman@nd.edu}
\emailAdd{bostdiek@nd.edu}

\abstract{
Extending the Standard Model by adding a scalar field transforming
as a septet under $SU(2)_L$ preserves the $\rho$ parameter at tree level and can
satisfy experimental constraints on the electroweak parameters $S$ and $T$.  
This work presents the first fully general phenomenological study of such an
extension.
We examine constraints on the septet model
couplings based on electroweak and Higgs observables, and use LHC searches for
new physics to bound the mass of the septet to be above $\sim 400$ GeV at a
$95\%$ CL.
}

\begin{document} 
\maketitle
\flushbottom

\numberwithin{equation}{section}
\section{Introduction}
\label{sec_Introduction}
The discovery of the Higgs boson begins a new era in particle physics. We now know many general details about the Higgs, although more precise measurements are needed. Many models of physics beyond the Standard Model (SM) contain more than one scalar which can contribute to breaking the electroweak symmetry. The amount that the scalars mix to create the physical particles can have a large impact on the couplings of the Higgs to both gauge bosons and fermions. Thus, measurements of the production cross section and decay channels of the Higgs offer new constraints on BSM models.

In addition to the new constraints provided by Higgs observables, 
models are still subject to the electroweak precision observables. 
The $\rho$ parameter ($\rho \equiv m_W^2/m_Z^2 \cos^2\theta_W$) measures the
ratio of the $W$ and $Z$ boson masses,
which depend on the properties of the Higgs sector. 
If the scalar sector is comprised only of singlets and doublets, the $\rho$
parameter is equal to one. However, a scalar field transforming as a triplet under $SU(2)_L$ does not preserve $\rho=1$ at tree level, so experimental constraints force a small vacuum expectation value (vev) of the triplet. For a review of exotic scalar fields see \cite{Kanemura:2013mc, Yagyu:2013kva}.
 
The general form of the $\rho$ parameter for an $SU(2)_L$ multiplet 
with weak isospin $j_{\phi}$ and hypercharge $Y_{\phi}$ is given by
 \begin{equation}
 \rho = \frac{\sum_{\phi}(j_{\phi}(j_{\phi}+1)-
 Y_{\phi}^2) v_{\phi}^2}{\sum_{\phi}(2Y_{\phi}^2) v_{\phi}^2}.
 \end{equation}

As $\rho$ is measured to be close to 1, the examination of models which protect the value
of $\rho$ at tree level is well motivated. After the doublet $(j=1/2,\; Y=
1/2)$, the next $SU(2)_L$ multiplet that leaves $\rho$ unity at tree level is
the septet $(j=3,\; Y=2)$. In fact, \cite{Hally:2012pu} showed that the doublet and 
the septet are the only multiplets which maintain $\rho =1$ and preserve 
perturbative unitarity of scattering amplitudes involving transverse 
W pairs and pairs of scalars. Interest in the septet has increased since the
discovery of the Higgs. Since the septet adds many charged scalars
to the Standard Model, it could be used to explain any discrepancy in
the observed $h \rightarrow \gamma \gamma$ rate. The authors of
\cite{Kanemura:2013mc, Hisano:2013sn} examined the production and decays of the
observed Higgs for the septet, while \cite{Killick:2013mya} explored how the
electroweak quantum numbers of an additional scalar field, such as the septet,
could be determined through a measurement of the Higgs-Higgs-vector-vector
coupling. The sum rules for general scalar representations of $SU(2)_L \times
U(1)_Y$ were studied in \cite{Grinstein:2013fia} using perturbative unitarity,
including the septet model as a specific example.

One artifact that arises in models containing a scalar with weak isospin $j>2$
is an accidental $U(1)$ symmetry at the renormalizable level.  If the scalar is 
to develop a vev to contribute to the breaking of the electroweak symmetry, the accidental $U(1)$ symmetry is broken as well. This accidental
symmetry cannot be spontaneously broken without generating a phenomenologically
unacceptable extra Nambu-Goldstone boson. All such models preserving the $U(1)$
are excluded by dark matter cosmological relic densities and direct-detection
cross section via Z exchange \cite{Earl:2013jsa}. In this paper we are not
concerned with dark matter and will thus explicitly break the accidental
symmetry with a non-renormalizable operator as done in \cite{Hisano:2013sn}.

This work presents the first constraints on a fully generic septet model with a
survey of the parameter space of all possible couplings. Earlier works used
simplified scenarios, ignoring certain parameters in the potential. We use the
full potential and explicitly calculate electroweak parameters and Higgs
observables from the masses and mixings between the doublet and the septet.
Using these calculations and the most current experimental results from 
ATLAS and CMS searches for new physics, we are able to place a general
bound on the mass of the septet.

The organization of the paper is as follows. In Section \ref{sec_Model} we
declare our notation and introduce the model Lagrangian and parameters. Section
\ref{section_Constraints} examines constraints on the septet model from 
Higgs observables and the $S$ and $T$ parameters. Using LHC searches for new physics, we
place bounds on the mass of the septet in Section \ref{sec_LHC}. Section
\ref{sec_conclusion} presents our conclusions.

\section{The Model}
\label{sec_Model}
In this section we present an overview of the septet model. 
This model includes the familiar scalar doublet field, $\Phi$, 
with quantum numbers $(1/2,\; 1/2)$ under $SU(2)_L\times U(1)_Y$. 
In addition, there is a second scalar field, $\chi$, with quantum numbers
$(3,2)$. This is a 7-dimensional representation of $SU(2)$, hence the
designation ``septet.'' These two scalar fields can be explicitly represented as
\begin{equation}
\Phi = \begin{pmatrix} \phi^+ \\ \phi^0 \end{pmatrix}, 
~~~ \chi = \begin{pmatrix} \chi^{+5} \\ \chi^{+4} \\ 
\chi^{+3} \\ \chi^{+2} \\ \chi^{+}_1 \\ \chi^{0} \\ \chi_2^{-}
\end{pmatrix}.
\label{eqn_parts}
\end{equation}
It is important to note that $\chi_1^+$ is not the antiparticle of $\chi_2^-$. 
When the neutral components of the scalar fields develop vevs, $\langle \phi^0
\rangle \equiv v_2$ 
and $\langle \chi^0 \rangle \equiv v_7$, mass is given to the $W$ and $Z$
according to the covariant derivatives:
\begin{equation}
m_W^2 = \frac{g^2}{2}(v_2^2 + 16 v_7^2), ~~~~~ m_Z^2 = \frac{g^2 + g^{\prime~2}}{2}(v_2^2 + 16 v_7^2).
\end{equation}
 At tree level, the vev of the septet does not alter the mass of the gauge bosons as long as 
\begin{equation}
v^2 = (174 \text{ GeV})^2 = v_2^2 + 16 v_7^2, ~~~~~v_2 = v \sin\beta, ~~ v_7 = \frac{1}{4} v \cos\beta.
\end{equation}
 To remain gauge invariant, the septet field cannot couple to fermions with
 renormalizable operators. It can also only couple to the doublet through quartic
 interactions containing each field multiplied by its conjugate. The most general
 potential is easiest to see using tensor methods \cite{Hisano:2013sn} and is given by
\begin{equation}
	\begin{aligned}
	V &= m_1^2 \Phi^2 + m_2^2 \chi^2 + \lambda (\Phi^{\dagger}\Phi)^2 
	-\frac{1}{\Lambda^3} \{(\chi^* \Phi^5 \Phi^*)+\text{H.C.}\}  \\
& + \sum_{A=1}^4 \lambda_A (\chi^{\dagger}\chi\chi^{\dagger}\chi)_A
+\sum_{B=1}^2 \kappa_B (\Phi^{\dagger}\Phi \chi^{\dagger}\chi)_B  + \frac{1}{\Lambda^2}\sum_{C=1}^{3} \eta_C (\Phi^{\dagger~2}\Phi^2 \chi^{\dagger}\chi)_C.
	\end{aligned}
	\label{eqn_Potential1}
\end{equation}
The tensor structure of the potential, with explicit forms for the
terms labeled with indices $A$, $B$, and $C$ is shown in Appendix \ref{appendix_1}.
The $\lambda_A$ term is a sum over four sub-terms with
slightly different tensor structures that collectively give rise to quartic
self-interactions of the septet field.  Similarly, the $\kappa_B$ term in \eqref{eqn_Potential1} is
a sum over two sub-terms; in this case allowing 4-point 
interactions between the septet and the doublet. Finally, 
the $\eta_C$ term contains three different tensor configurations of a dimension 
six operator which affects the mass relations of the septet. The dimension-6 $\eta_C$ term
is necessary to maintain consistency of the effective field theory, since we are
including the dimension-7 term to break the accidental $U(1)$ symmmetry. There
are no possible dimension-5 terms.

It will be shown in Section \ref{section_Constraints} that the LHC Higgs observations force a small value of $v_7$. Since
$v_7$ must be small, the septet quartic couplings $\lambda_A$ will
contribute negligibly to the mass of the septet, and the septet three-point interactions
will be small. The exact values of the $\lambda_A$ will thus not have a large
effect on this study, as very precise measurements would be needed to determine
these values. Another result of $v_7$ being small is that the possible
dimension-6 operators other than $\eta_C$ either have negligible effects or can
be absorbed into $\lambda_A$ or $\kappa_B$.

Keeping in mind that we will later show $v_7$ to be small, we here show the
masses of the septet particles, neglecting the contribution from $v_7$ and
keeping the contributions from $m_2$ and $v_2$.
\begin{eqnarray}
m_{\chi^{+5}}^2 &=& m_2^2 + v_2^2 (\kappa_1+ \eta_1 \frac{v_2^2}{\Lambda^2}) \nonumber \\
m_{\chi^{+4}}^2 &=& m_2^2 + v_2^2 (\kappa_1+ \eta_1 \frac{ v_2^2}{\Lambda^2})+ v_2^2\frac{1}{6}(\kappa_2+\eta_2\frac{ v_2^2}{\Lambda^2}) \nonumber \\
m_{\chi^{+3}}^2 &=& m_2^2 + v_2^2 (\kappa_1+ \eta_1 \frac{ v_2^2}{\Lambda^2})+ v_2^2\frac{1}{3}(\kappa_2+\eta_2\frac{ v_2^2}{\Lambda^2})+\eta_3 \frac{v_2^4}{15 \Lambda^2} \nonumber \\
m_{\chi^{++}}^2 &=& m_2^2 + v_2^2 (\kappa_1+ \eta_1 \frac{ v_2^2}{\Lambda^2})+ v_2^2\frac{1}{2}(\kappa_2+\eta_2\frac{ v_2^2}{\Lambda^2}) +\eta_3 \frac{v_2^4}{5 \Lambda^2}\label{eqn_SeptetMasses}\\
m_{\chi_1^{+}}^2 &=& m_2^2 + v_2^2 (\kappa_1+\eta_1 \frac{ v_2^2}{\Lambda^2})+ v_2^2\frac{2}{3}(\kappa_2+\eta_2\frac{ v_2^2}{\Lambda^2}) +\eta_3 \frac{2 v_2^4}{5 \Lambda^2}\nonumber \\
m_{\chi^0}^2 &=& m_2^2 + v_2^2 (\kappa_1+\eta_1  \frac{ v_2^2}{\Lambda^2})+ v_2^2\frac{5}{6}(\kappa_2+\eta_2\frac{ v_2^2}{\Lambda^2})+\eta_3 \frac{2 v_2^4}{3 \Lambda^2} \nonumber \\
m_{\chi_2^{+}}^2 &=& m_2^2 + v_2^2 (\kappa_1+ \eta_1 \frac{ v_2^2}{\Lambda^2})+ v_2^2(\kappa_2+\eta_2\frac{ v_2^2}{\Lambda^2})+\eta_3 \frac{v_2^4}{\Lambda^2} \nonumber
\end{eqnarray}
Equations \eqref{eqn_Potential1} and \eqref{eqn_SeptetMasses} show that the
parameters $\kappa_1$ and $\eta_1$ couple the septet and doublet equally for each
particle. However, the contributions from $\kappa_2$, $\eta_2$, and $\eta_3$ increase for lower
components of the septet representation. This trend will have a direct impact on
the $S$ and $T$ parameters as well as the $h\rightarrow \gamma \gamma$
coupling. We also see that $\eta_1$ ($\eta_2$) can be reabsorbed into $\kappa_1$ ($\kappa_2$).

In order to not have the charged components of the septet develop a vev, we do
not allow for a negative $m_2^2$; all of the vev of the septet thus comes 
from the tadpole term of the dimension-7 operator in the first line 
of \eqref{eqn_Potential1}. The fact that this term contains only a single septet 
field allows for the tadpole and explicitly breaks the accidental $U(1)$ symmetry 
mentioned in the introduction, thereby preventing a massless Nambu-Goldstone boson. Reference
\cite{Hisano:2013sn} has a more detailed analysis of a possible UV completion
leading to this dimension-7 term. Momentarily
setting $\lambda_A$,  $\kappa_B$, and $\eta_C$ in \eqref{eqn_Potential1} to zero
gives a septet vev of
\begin{equation}
	v_7 = \frac{v_2^6}{\sqrt{6} m_2^2 \Lambda^3}.
\label{eqn_septetvev}
\end{equation}
As the mass of the septet is decoupled or $\Lambda$ becomes large, the vev of
the septet goes to zero and the Standard Model is recovered. From this point on,
we allow $\lambda_A$, $\kappa_B$, and $\eta_C$ to vary.

The septet contains two distinct singly charged particles as well as a neutral one. These will mix with the charged and neutral components of the doublet to form  physical particles. We define $\Phi_0 = v_2 + (\phi^0_R + \imath \phi^0_I)/\sqrt{2}$ and $\chi^0 = v_7 + (\chi^0_R + \imath \chi^0_I)/\sqrt{2}$ with the rotations given by
\begin{eqnarray}
\begin{pmatrix}\phi^0_R \\ \chi^0_R \end{pmatrix} &= &\begin{pmatrix} \cos\alpha & -\sin\alpha \\ \sin\alpha & \cos \alpha \end{pmatrix} \begin{pmatrix}h_0 \\ H_0\end{pmatrix}, \nonumber \\
\begin{pmatrix}\phi^0_I \\ \chi^0_I \end{pmatrix} &= &\begin{pmatrix}\sin \beta & -\cos\beta \\ \cos\beta & \sin\beta  \end{pmatrix} \begin{pmatrix}G_0 \\ A_0\end{pmatrix} , \\
\begin{pmatrix}\phi^+ \\ \chi_1^+ \\ \chi_2^+ \end{pmatrix} &= &S_{Charge} \begin{pmatrix} G^+ \\ H_1^+ \\ H_2^+ \end{pmatrix} . \nonumber
\end{eqnarray}
where $h_0$ is the lightest neutral CP even Higgs and the one observed at the LHC, 
$H_0$ is the heavier neutral CP even Higgs, $A_0$ is the neutral CP odd Higgs,
and $H_1^+$ $(H_2^+)$ is the lighter (heavier) singly charged Higgs. 
$G_0$ and $G^+$ are the goldstones eaten by the $W$ and $Z$.


\section{Constraints from observables}
\label{section_Constraints}

To study the phenomenology of this model we first examine the couplings of
the Higgses. The new couplings between the CP-even Higgses and
vector bosons are 
\begin{equation}
	\begin{aligned}
g_{hVV} = \frac{g^2 v}{\sqrt{2}} (4\cos\beta \sin\alpha + \cos\alpha \sin\beta) = 
g_{hVV}^{SM}(4\cos\beta \sin\alpha + \cos\alpha \sin\beta) , \\
g_{HVV} = \frac{g^2 v}{\sqrt{2}} (4\cos\alpha \cos\beta -\sin\alpha\sin\beta) = g_{hVV}^{SM} (4\cos\alpha \cos\beta -\sin\alpha\sin\beta).
\label{eqn_hVV}
\end{aligned}
\end{equation}
Of course, the septet field cannot couple directly to Standard Model fermions. 
So the Higgs-fermion coupling is only through the doublet:
\begin{equation}
	\begin{aligned}
g_{hf\overline{f}} = \frac{y_f}{\sqrt{2}} \cos\alpha = \frac{m_f}{v_2}
\cos\alpha = g_{hf\overline{f}}^{SM} \frac{\cos\alpha}{\sin\beta}, \\
g_{Hf\overline{f}} = -\frac{y_f}{\sqrt{2}} \sin\alpha = -\frac{m_f}{v_2}
\sin\alpha = -g_{hf\overline{f}}^{SM} \frac{\sin\alpha}{\sin\beta}.
\label{eqn_hff}
\end{aligned}
\end{equation}
From \eqref{eqn_hVV} and \eqref{eqn_hff}, note that the SM is recovered as $\alpha \rightarrow 0$ and $\beta \rightarrow \pi/2$.

\begin{table}[t]
\begin{center}
\begin{tabular}{|ccc|}
\hline
Signal	& ATLAS	&	CMS\\
\hline
$\mu_{WW}$ & $1.25 \pm 0.43$ & $0.68 \pm 0.20$ \\
$\mu_{ZZ}$ & $1.20 \pm 0.58$ & $0.92\pm 0.28$ \\ 
$\mu_{\gamma\gamma}$ & $1.76\pm0.50$ & $0.77\pm0.27$ \\
$\mu_{bb}$ & $0.47\pm2.17$ & $1.15\pm0.62$ \\
$\mu_{\tau\tau}$ & $0.44\pm1.55$ & $1.10\pm0.41$ \\
\hline
\end{tabular}
\end{center}
\caption{Higgs signal strengths from ATLAS \cite{Aad:2012tfa} and CMS \cite{Chatrchyan:2013lba}. The signal stength $\mu_{xx}$ is
defined as the bound on the Higgs decay rate $\Gamma(h\rightarrow x x)$ divided by the SM expectation.}
\label{tbl_HiggsStrengths}
\end{table}%

\begin{figure}[t]
	\centering
        	\begin{subfigure}[b]{0.465\textwidth}
                \includegraphics[width=\textwidth]{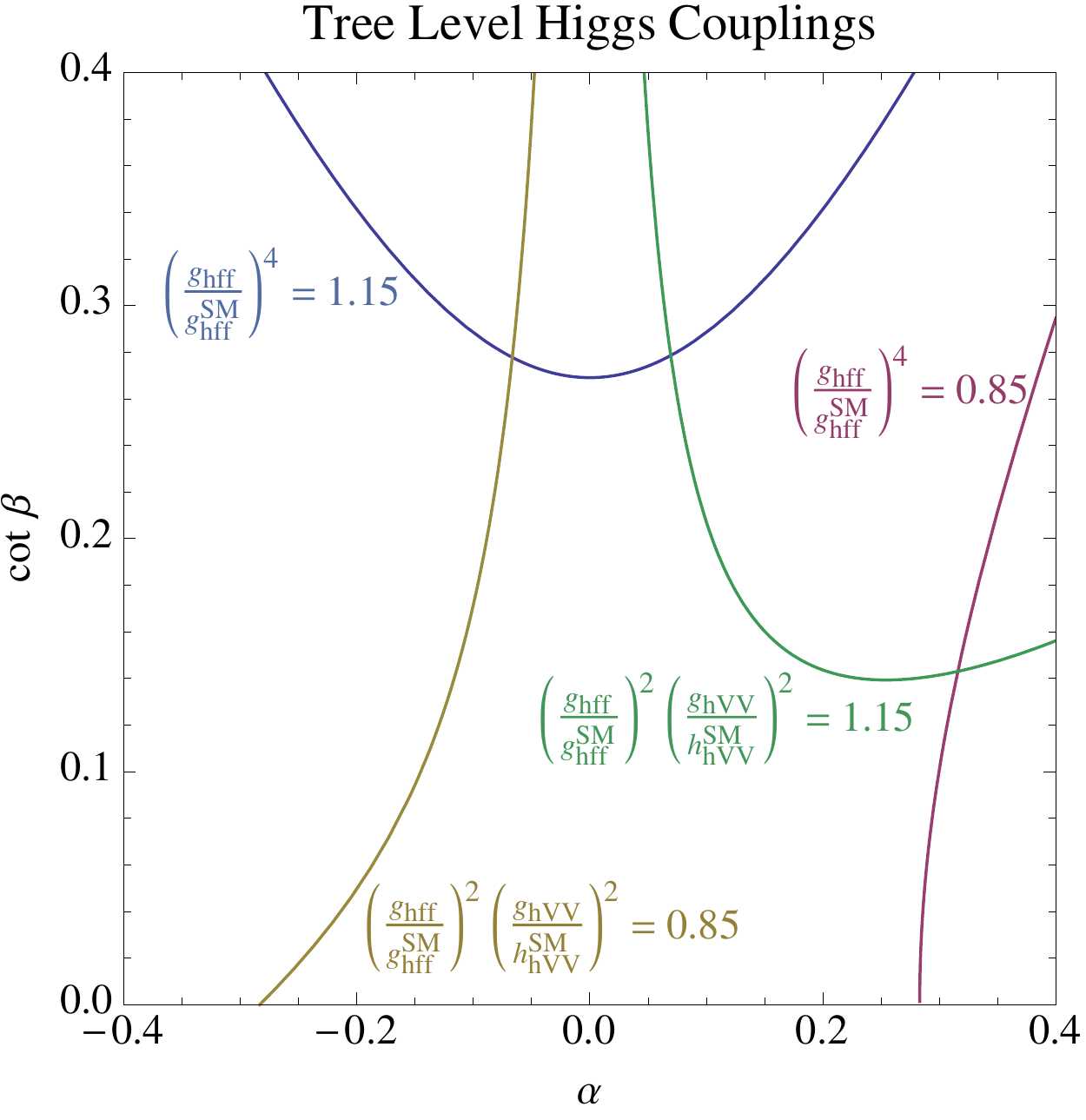}
          \end{subfigure}
          ~~
          \begin{subfigure}[b]{0.45\textwidth}
                \includegraphics[width=\textwidth]{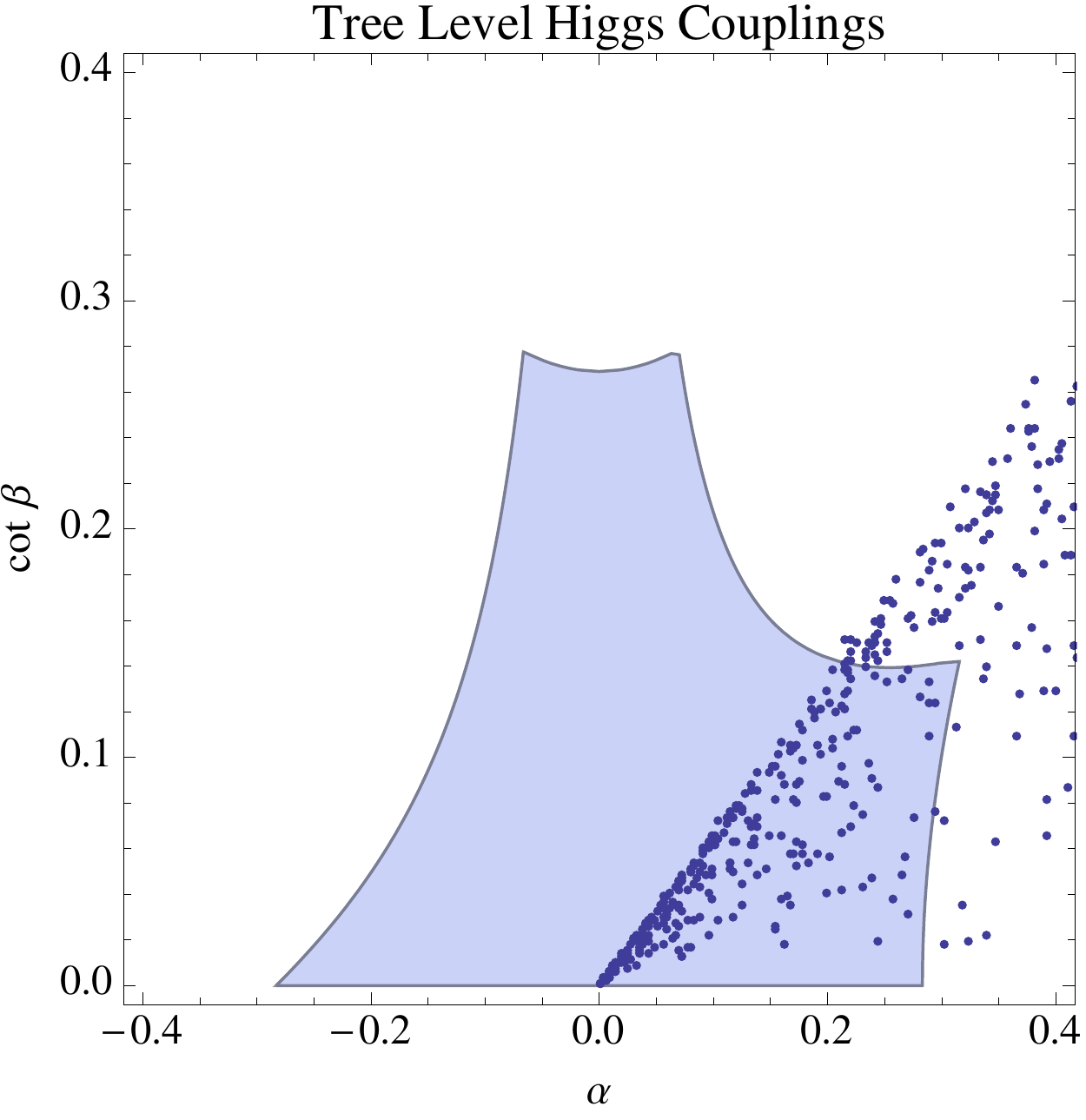}
          \end{subfigure}
         \caption{The left panel displays the constraints coming
		 from the tree level Higgs signal strengths. The blue (red) line enforces 
		 that gluon fusion followed by a decay to fermions is not too
		 large (small). This is the constraint from \eqref{HC1}. The yellow (green) 
		 lines are for gluon fusion followed by decays to $W$ and $Z$
		 bosons, which is the constraint shown in \eqref{HC2}. These constraints set the
		 area of allowed parameter space in $\alpha$ and $\cot\beta$, but the septet model
		 cannot reproduce the observed Higgs mass throughout this entire region. The right
		 panel displays model points chosen with random, perturbative couplings with a 
		 minimized potential and $m_h=125.5$ GeV according to equation \eqref{eqn_modelChoice}.
		 The shaded area is allowed by the bounds shown in the left panel. 
		 To pass the tree level Higgs couplings and generate the correct Higgs mass, the model points must have $\cot\beta \lesssim 0.14$.}
\label{fig_TreeLevelHiggs}
\end{figure}

Some bounds on the Higgs signal strengths for ATLAS  \cite{Aad:2012tfa} and
CMS \cite{Chatrchyan:2013lba} are shown in Table \ref{tbl_HiggsStrengths}.
The signal strengths are the observed rates divided by the SM expectation.
In these categories, one experiment measures 
a value above the SM expectation and the other experiment a value below the SM
expectation. We make the conservative 
assumption that the tree level processes will not deviate from the SM by more
than 15 percent. Using this assumption, we make the following cut for gluon
fusion with decays to fermions:
\begin{equation}
	0.85  \le \left(g_{hf\overline{f}}/g_{hf\overline{f}}^{SM}\right)^4 \le 1.15  .
	\label{HC1}
\end{equation}
A similar cut is made for gluon fusion followed by decays to gauge bosons:
\begin{equation}
	0.85 \le \left( \left(g_{hf\overline{f}}
	g_{hVV}\right)/\left(g_{hf\overline{f}}^{SM} g_{hVV}^{SM}\right)
	\right)^2 \le 1.15 .
	\label{HC2}
\end{equation}
The left panel of Figure \ref{fig_TreeLevelHiggs} shows the effect of the fermion decay constraints in blue and red, while the gauge boson decays are constrained in green and yellow.

These constraints have come only from the production and decay of the Higgs. These couplings compared to the SM are only determined by the mixing angles of the neutral Higgses and the vevs, with no other assumptions about the masses or couplings in the model. To further constrain the model, we generate random model points which pass constraints \eqref{HC1} and \eqref{HC2}, yield the observed Higgs mass,  do not allow for stable charged scalars, and have perturbative couplings. The model points chosen then have the following form.
\begin{equation}
-2 \le \{\lambda_1, \lambda_2, \lambda_3, \lambda_4, \kappa_1, \kappa_2,\eta_3\} \le 2,~~~~~0\le\cot\beta\le0.3, ~~~\text{and}~~~ 0\le m_2 \le 2 \text{ TeV.} 
\label{eqn_modelChoice}
\end{equation} 
At each point, the values of $m_1$ and $\Lambda$
are used to minimize the potential and $\lambda$ sets $m_{h_0}=125.5 \text{ GeV}$. 
The neutral Higgs rotation angle, $\alpha$ is an output of the model. 
We only keep points which generate $\alpha$ reproducing the allowed Higgs couplings. 
To satisfy charged dark matter bounds, we assume the mass hierarchy
\begin{equation}
	m_{\chi^{5+}} > m_{\chi^{4+}} > m_{\chi^{3+}} > m_{\chi^{2+}}.
\label{mass_hierarchy}
\end{equation}
This and the mass relations in \eqref{eqn_SeptetMasses} lead to the assumption that $\kappa_2 \le 0$. It is possible that three body (or higher)
decays could be used to satisfy dark matter bounds if this mass hierarchy is not used. Relaxing this assumption would require an in-depth analysis of actual
decay widths and possible charge injection into the early universe. While this is beyond the scope of this paper, a detailed examination of $\kappa_2 >0$ would be a rich topic of study.

In the right panel of Figure \ref{fig_TreeLevelHiggs} the model points are plotted in the same window as 
the $(\alpha, \cot\beta)$ tree-level Higgs coupling constraints. It is not possible to generate the observed 
Higgs mass for all values of $(\alpha, \cot\beta)$. The constraint that the gauge boson signal strengths are 
not greater than the SM expectation by more than $15\%$ forces $\cot\beta \lesssim 0.14$, or
\begin{equation}
v_7 < 6 \text { GeV}.
\label{eqn_v7less6}
\end{equation}
This small value of $v_7$ justifies dropping $\mathcal{O}(g^2 v^2_7)$ contributions to the masses 
given in \eqref{eqn_SeptetMasses}.

\subsection{$S$ and $T$ Parameters}

\begin{figure}[t]
        \centering
        \begin{subfigure}[b]{0.465\textwidth}
                \includegraphics[width=\textwidth]{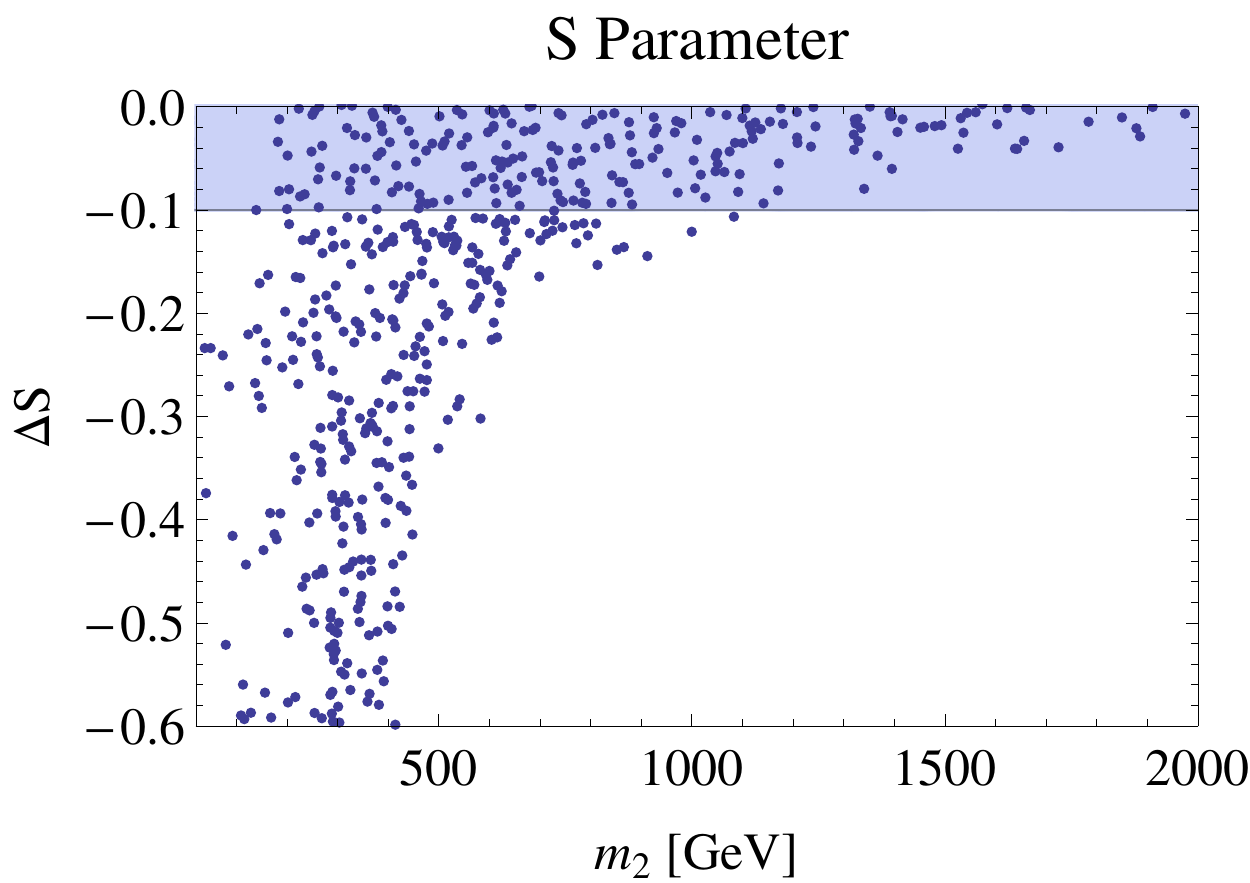}
        \end{subfigure}%
        ~~ 
        \begin{subfigure}[b]{0.45\textwidth}
                \includegraphics[width=\textwidth]{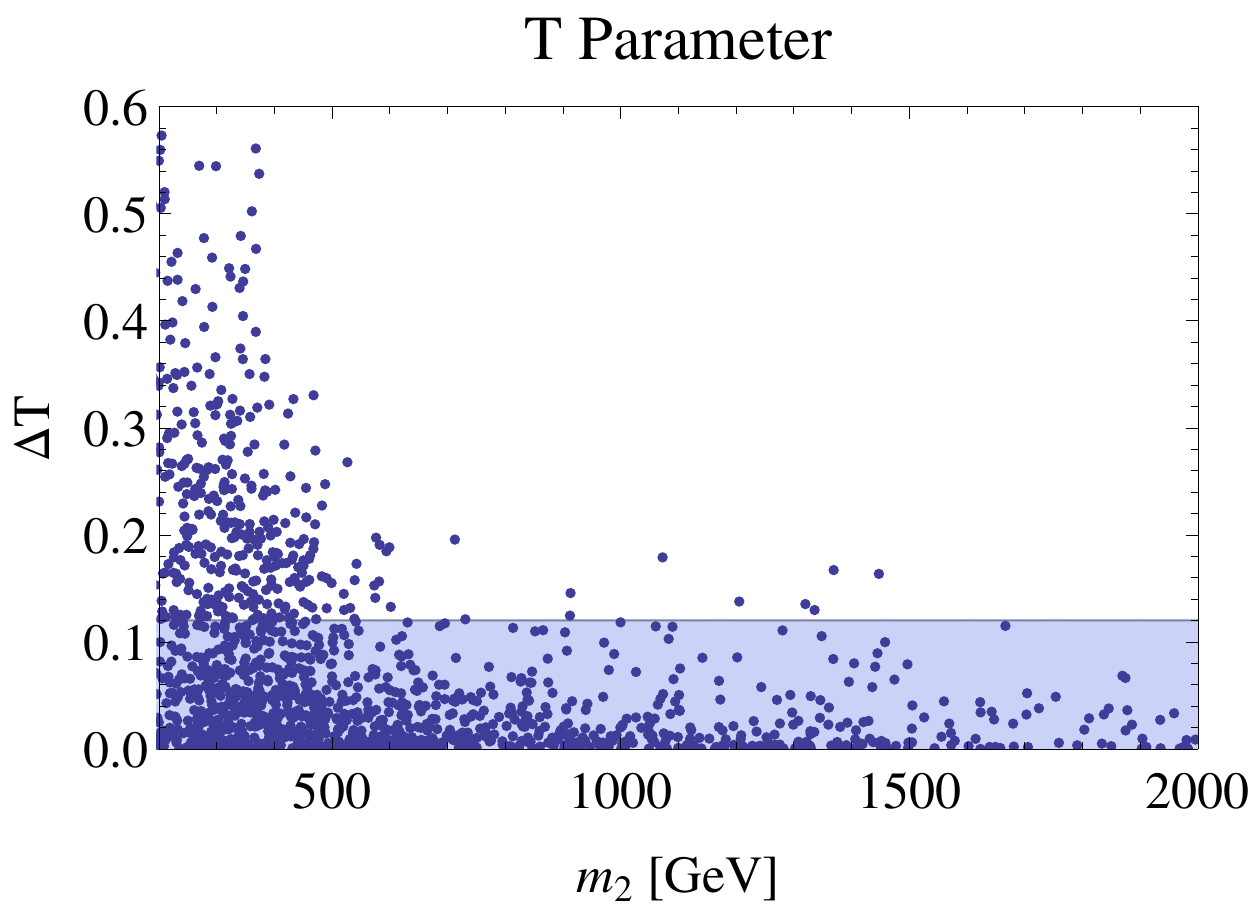}
        \end{subfigure}%
	\caption{The left (right) panel shows the septet contributions to the
		$S$ ($T$) parameter plotted against the septet mass parameter
		$m_2$. The model points are randomly chosen according to 
		equations \eqref{HC1}, \eqref{HC2}, and \eqref{eqn_modelChoice}
		such that the observed Higgs mass is generated and the tree-level 
		Higgs couplings do not deviate by more than $15\%$ from the SM value.
		The shaded area is the allowed region for $S$ and $T$.
		At large values of $m_2$ the contributions to these parameters 
		are small. However, there are still many model 
		points with small values of $S$ and $T$ and low values of
		 $m_2$. At low masses the $S$ and $T$ parameters are 
		controlled by $\kappa_2$ and $\eta_3$, so these parameters are
		forced to take values close to zero.}
        \label{fig_SandT_1}
\end{figure}

One of the main motivations of the septet model is that it does not affect the $T$ parameter at tree level ($\alpha T=\rho-1$). However, it can induce changes at loop level.  The scalar contributions to the vacuum polarization at one loop are given by
\begin{equation}
	\begin{aligned}
	\Pi^{AB}(q^2)  = \frac{T^A_{ij}T^B_{ji}}{16\pi^2} \bigg[ (m_i^2& +
		m_j^2)(1+\xi) -
	2\,f_2(m_i,m_j) + 2\, m_i^2 (\log m_i^2 - \xi -1)  \;  \\
	&  + q^2\left( 2\, f_1(m_i,m_j) - \frac{\xi}{3}\right) +
\mathcal{O}(q^4)\bigg],
	\end{aligned}
\end{equation}
where
\begin{eqnarray}
f_1(m_i,m_j) &=& \int_0^1 dx~ x(1-x) \log(x m_i^2 + (1-x)m_j^2) \nonumber\; , \\
f_2(m_i, m_j) &=& \int_0^1 dx~ (x m_i^2+(1-x)m_j^2)\log(xm_i^2 +(1-x) m_j^2)
\nonumber \; , \\
\xi &=& \lim_{\epsilon \rightarrow 0}~ 2/\epsilon - \gamma + \log 4\pi. \nonumber
\end{eqnarray}
The masses, $m_{(i,j)}$, are given in \eqref{eqn_SeptetMasses} and the $SU(2)_L$
generators for the septet, $T^A_{ij}$, are given in Appendix \ref{section_SU(2)Gen}. The $S$ and $T$ parameters are then
\begin{eqnarray}
S &=& 16\pi \left[\Pi^{\prime~ 3 3}(0) - \Pi^{\prime~ 3Q}(0)\right],  \\
T &=& \frac{4\pi}{s_W^2 c_W^2 m_Z^2}\left[\Pi^{11}(0) - \Pi^{33}(0)\right].
\end{eqnarray}
Note that $\xi$ cancels in the calculation so that $S$ and $T$ are finite.
The oblique parameters are larger when the mass of different parts of a multiplet
are non-degenerate. As $v_7$ is small from the tree level constraints
\eqref{eqn_v7less6}, the mass splitting of the septet comes from $\kappa_2$ and
$\eta_3$, although the $\eta_3$ contribution is suppressed by an extra factor of
$(v_2/\Lambda)^2$.
The contributions to $S$ and $T$ should go to zero as the septet is decoupled.
The randomly generated model points contributions to $S$ and $T$ are plotted against the common mass parameter of the septet, $m_2$, in figure \ref{fig_SandT_1}. The shaded box shows the allowed regions \cite{Baak:2012kk} where
\begin{equation}
|\Delta S| \le 0.12 ~~~\text{and}~~~ |\Delta T| \le 0.10.
\label{eqn_S&TConstraints}
\end{equation}

The septet protects the $T$ parameter at tree level, and the loop contributions
are smaller for $T$ than $S$. For points with $m_2 \gtrsim 600 \text{ GeV}$ the
$T$ parameter is small whereas the $S$ parameter is not necessarily small until
$m_2 \gtrsim 1 \text{ TeV}$.

\begin{figure}[t]
        \centering
        \begin{subfigure}[b]{0.465\textwidth}
                \includegraphics[width=\textwidth]{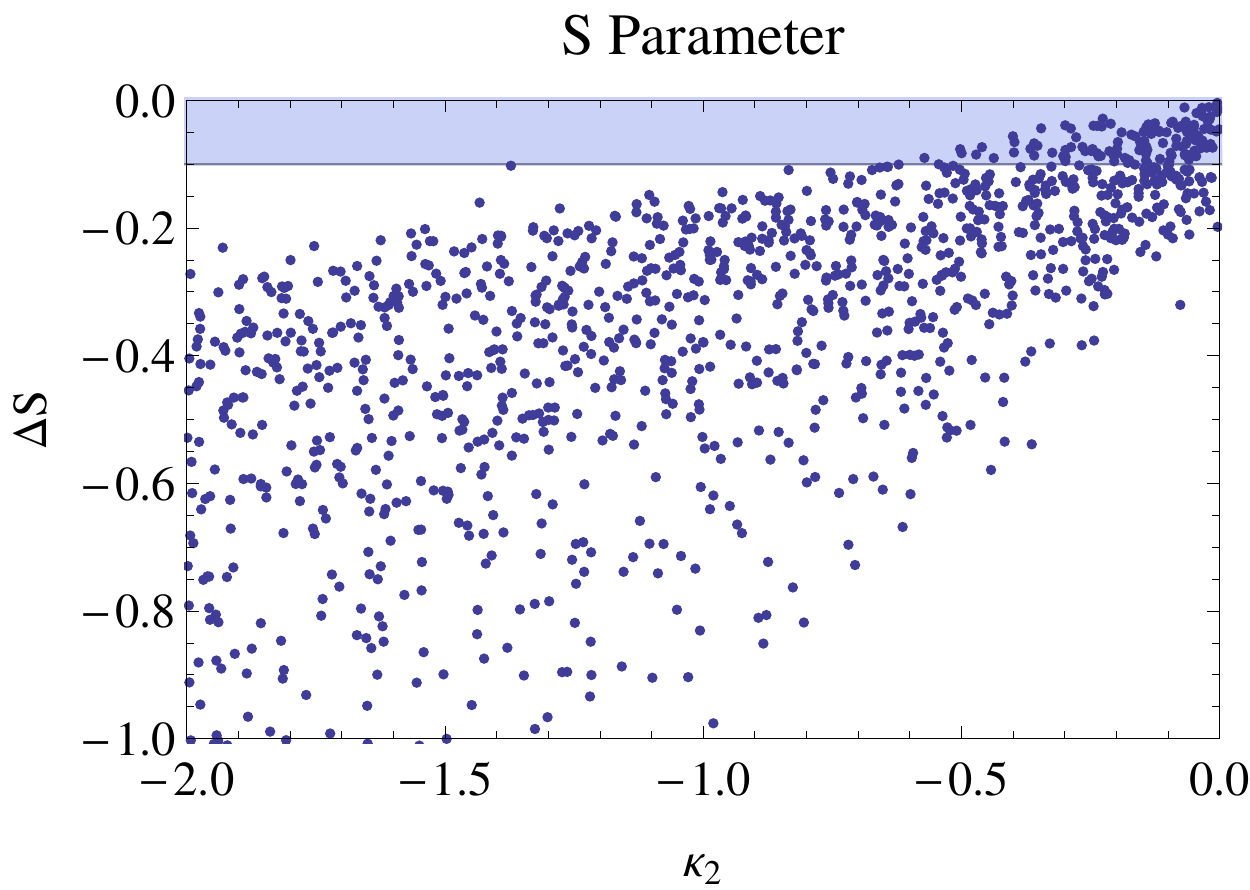}
        \end{subfigure}%
        ~~ 
        \begin{subfigure}[b]{0.45\textwidth}
                \includegraphics[width=\textwidth]{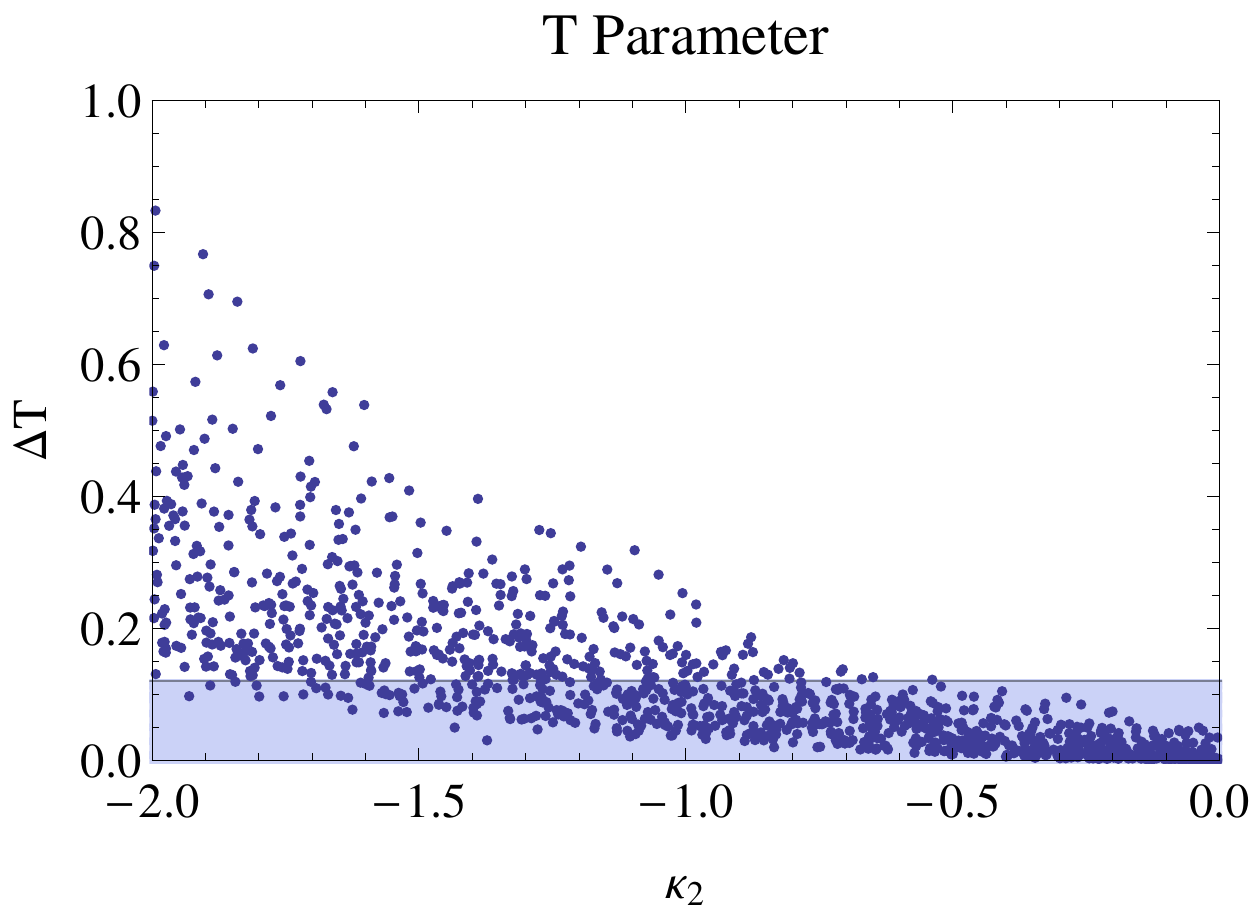}
        \end{subfigure}%
	\caption{The left (right) panel shows the septet contributions to the
		$S$ ($T$) parameter plotted against $\kappa_2$ which couples the septet to
		the doublet in the potential shown in equation \eqref{eqn_Potential1}. The model
		points are randomly chosen according to 
		equations \eqref{HC1}, \eqref{HC2}, and \eqref{eqn_modelChoice}
		such that the observed Higgs mass is generated and the tree-level 
		Higgs couplings do not deviate by more than $15\%$ from the SM value.
		In addition, the model points in this figure have $m_2 < 500$ GeV to allow
		for large contributions to $S$ and $T$. The coupling $\kappa_2$ alters the mass
		of the septet differently for each component as seen in equation
		\eqref{eqn_SeptetMasses}. Large absolute values of $\kappa_2$ lead to greater
		mass splitting and thus more contribution to $S$ and $T$.}
        \label{fig_SandT_2}
\end{figure}

Figure \ref{fig_SandT_2} shows the same model points as before with the added
condition that $m_2 < 500$ GeV, 
such that the contributions to $S$ and $T$ can be substantial. 
If $\kappa_2$ is large, the mass splitting among different components of the septet
is large, leading to greater contributions to $S$ and $T$. The mass hierarchy chosen so as to not have a relic abundance 
of charged septet states has already forced $\kappa_2 \le 0$ and now the electroweak observables force small values 
of $\kappa_2$ when the mass parameter $m_2$ is small.

\subsection{Higgs to $\gamma \gamma$}

The septet model contains many more charged particles coupling to the Higgs than
does the SM. The experimental observations of the rate of $h\rightarrow \gamma
\gamma$ leave room for non-SM-like behavior. However,
CMS measures the rate to be below the SM rate whereas ATLAS measures it to be
above, as shown in Table \ref{tbl_HiggsStrengths}. With this in mind, we allow a greater range for this signal 
than we did for the gluon fusion signals by only requiring 
\begin{equation}
0.5 \le \Gamma(h\rightarrow \gamma \gamma)/\Gamma(h\rightarrow \gamma \gamma)_{SM} \le 2.0.
\label{eqn_hggcut}
\end{equation}

\begin{figure}[t]
        \centering
       \begin{subfigure}[b]{0.45\textwidth}
                \includegraphics[width=\textwidth]{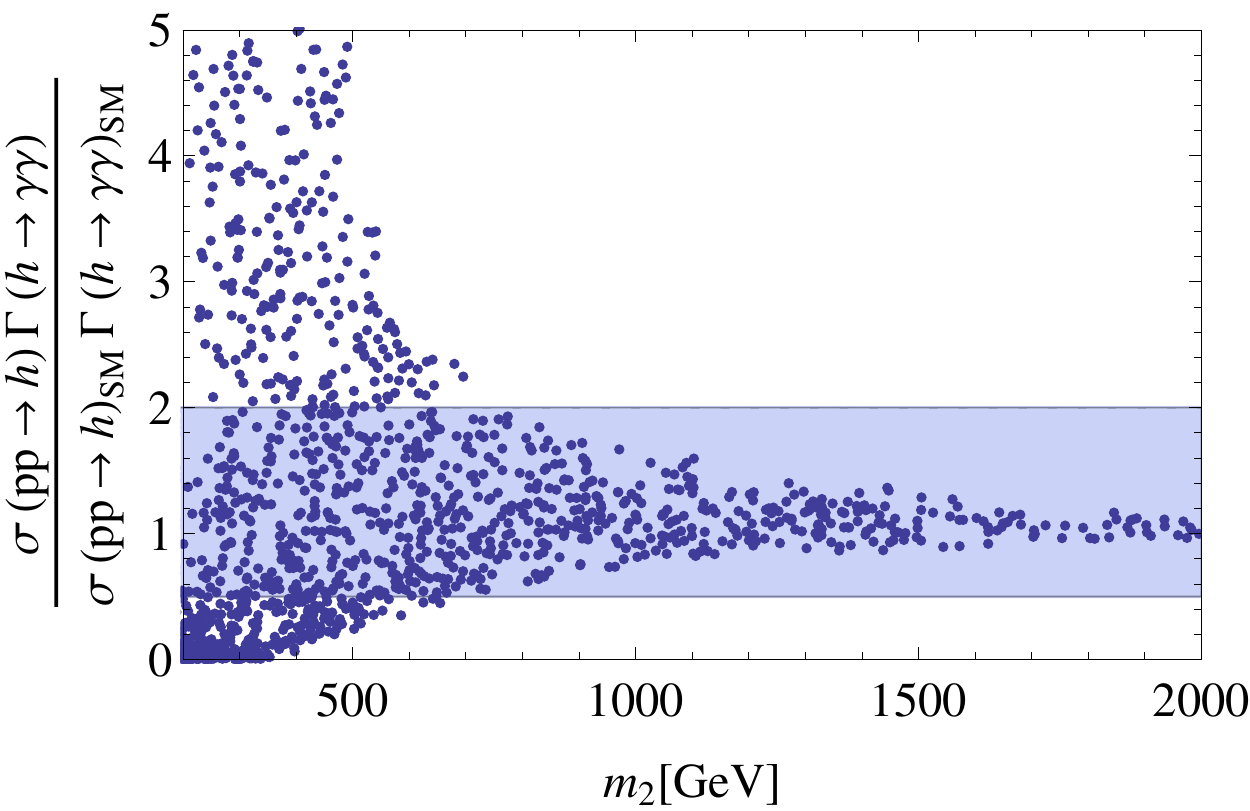}
        \end{subfigure}%
        ~~ 
        \begin{subfigure}[b]{0.45\textwidth}
                \includegraphics[width=\textwidth]{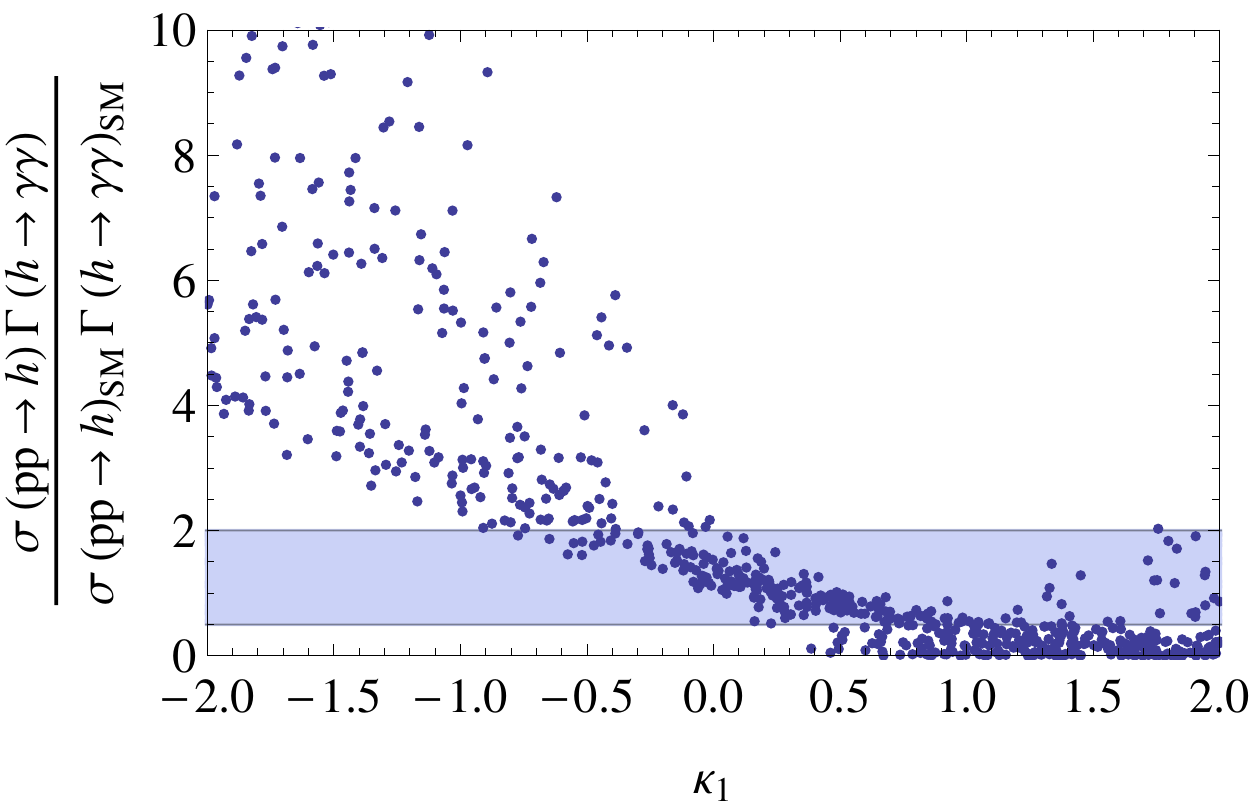}
        \end{subfigure}%
	\caption{The left panel plots the $h\rightarrow\gamma\gamma$
		signal strength of the model points against the septet mass. 
		The model points are randomly chosen according to 
		equations \eqref{HC1}, \eqref{HC2}, and \eqref{eqn_modelChoice}
		such that the observed Higgs mass is generated and the tree-level 
		Higgs couplings do not deviate by more than $15\%$ from the SM value. 
		The shaded area corresponds to the allowed region. At large values of $m_2$ the 
		septet contributions to the $h\rightarrow \gamma \gamma$ signal are small.
		The coupling of the Higgs to the charged septet fields allowing the di-photon decay
		comes from the $\kappa_1$, $\kappa_2$, and $\eta_3$ terms.
		At low masses the $S$ and $T$ parameters force small 
		values for $\kappa_2$ and $\eta_3$. 
		Because of this, $\Gamma(h\rightarrow\gamma\gamma)$
		is affected most by the remaining coupling of the doublet to the septet,
		$\kappa_1$. The right panel shows the signal strength plotted against
		 $\kappa_1$, for small septet masses ($m_2 <500$ GeV). A negative value 
		of $\kappa_1$ leads to an increase in $\Gamma(h\rightarrow\gamma\gamma)$ while a 
		positive value decreases $\Gamma(h\rightarrow\gamma\gamma)$.}
        \label{fig_hgammagamma}
\end{figure}

The partial width of $h\rightarrow\gamma\gamma$ is given by \cite{Carena:2012xa} as
\begin{equation}
\Gamma(h\rightarrow\gamma\gamma) = \frac{\alpha^2 m_h^3}{512 \pi^3} 
\left|\frac{g_{hVV}}{m_V^2} Q_V^2 A_1(\tau_V) + 
\frac{2 g_{hf\overline{f}}}{m_f} N_{c,f} Q^2_f A_{1/2}(\tau_f) + 
N_{c,s} Q_S^2 \frac{g_{hSS}}{m_S^2} A_0(\tau_S)\right|^2 ,
\label{eqn_hgammagamma1}
\end{equation}
where $v=174$ GeV, $Q$ is the charge of the particle, $N_c$ is the 
number of colors, $\tau_x = 4 m_x^2 / m_h^2$, and the $A_S$ functions are given by
\begin{eqnarray}
A_1(x) &=& -x^2 \left(2 x^{-2} + 3 x^{-1} + 3(2x^{-1}-1) f(x^{-1}) \right), \nonumber \\
A_{1/2}(x) &=& 2 x^2 \left(x^{-1} + (x^{-1}-1)f(x^{-1})\right), \nonumber \\
A_{0}(x) &=& -x^2 \left(x^{-1}-f(x^{-1})\right), \nonumber \\
f(x) &=& \arcsin^2(\sqrt{x}).
\label{eqn_hgg1loop}
\end{eqnarray}
For the $W$ and top quark, $A_1(\tau_W) = -8.3$ and $A_{1/2}(\tau_t) = 1.4$. If the scalar mass is greater than the Higgs mass
$A_0(\tau_S) \sim 1/3$. Thus the scalar contributions
work against the SM dominant $W$ boson contributions if the $g_{hSS}$ couplings are positive. 
This allows the septet to decrease and potentially flip the sign of the
$g_{h\gamma\gamma}$ coupling with positive Higgs-septet couplings. If the
couplings are negative then the contributions increase the partial decay width.
Figure \ref{fig_hgammagamma} shows the random model points
passing the tree-level coupling constraints. Again, the blue shaded region is
the allowed region defined by \eqref{eqn_hggcut}. As the septet decouples the SM
expectation is recovered as shown in the left panel. 
The tree-level Higgs couplings force a small value of the neutral mixing angle, $\alpha$.
This implies that the observed Higgs
is mostly doublet like, therefore the couplings $g_{hSS}$ come from the $\kappa_1$,
$\kappa_2$, and $\eta_3$ terms. At low masses, $S$ and $T$ 
force small $\kappa_2$ and $\eta_3$. Thus the common coupling of $h_0$ to the septet, 
$\kappa_1$, has the most effect on the partial decay width to two photons. 
A negative value of $\kappa_1$ increases $\Gamma(h\rightarrow\gamma\gamma)$ and a positive value decreases it. The right panel shows this trend, where the only model
points shown have $m_2 < 500$ GeV. As $\kappa_1$ does not affect $S$ and $T$, any future precise measurement of $h\rightarrow\gamma\gamma$ can be accommodated by the septet model.

\section{Bounds from the LHC}
\label{sec_LHC}
After taking all of the constraints from Section \ref{section_Constraints} into
account, the allowed parameter space contains $\cot \beta < 0.14$ and $\alpha <
0.25$. The constraints leading to these bounds all came from electroweak
observables or properties of the observed Higgs. The question still remains: how
do direct searches for new particles at the LHC constrain the septet model?  To answer this
question, we create a grid of model points with $0\le \cot\beta \le 0.14$ and
$100\le m_2\le 700$ GeV, with the rest of the model parameters following
\eqref{eqn_modelChoice}.  The particular points are chosen to sample the
parameter space at very small $\cot\beta$ 
and provide good coverage for $m_2$ between 300 and 500 GeV, where the inclusive cross section ranges between 
hundreds of fb to tens of fb. The grid values are given by
\begin{equation}
\begin{aligned}
m_2 &= \{100, 200, 300, 350, 400, 450, 500, 600, 700\} \text{ GeV} \\ 
\cot\beta&=\{0.001, 0.005, 0.01, 0.02, 0.04, 0.06, 0.08, 0.10, 0.12, 0.14\}.
\end{aligned}
\end{equation}
Each grid point is also forced to follow all of the constraints discussed in Section \ref{section_Constraints}.

As previously noted, the septet model offers a rich landscape of multiply
charged particles that can be phenomenologically useful. 
However, this relatively large number of particles
makes a direct search method difficult. For our search, we implemented the
septet model in FeynRules \cite{Alloul:2013bka}. The events were then simulated
with MadGraph 5 \cite{Alwall:2011uj}, using Pythia 6.4 \cite{Sjostrand:2006za}
for hadronization and PGS \cite{pgs} for detector simulation.

Using these tools, we considered the production of any new particle pair at the LHC with $\sqrt{s}=8$
TeV. As the septet does not couple directly to SM fermions or gluons, the
production mechanism must be electroweak, going through either a neutral or
charged current Drell-Yan type process. It is also possible to create the new
particles through VBF leading to four-point interactions with two gauge bosons
and two new scalars. The cross section for the VBF process is less than
$10\%$ of the total new scalar cross section. The presence of so many new states
makes the VBF simulations extremely computationally demanding, so we do not include them in
this analysis. Our results are thus conservative, but a dedicated search could
find the VBF process useful.

\begin{figure}[t]
        \centering
        \begin{subfigure}[b]{0.445\textwidth}
      	  \includegraphics[width=.925\textwidth]{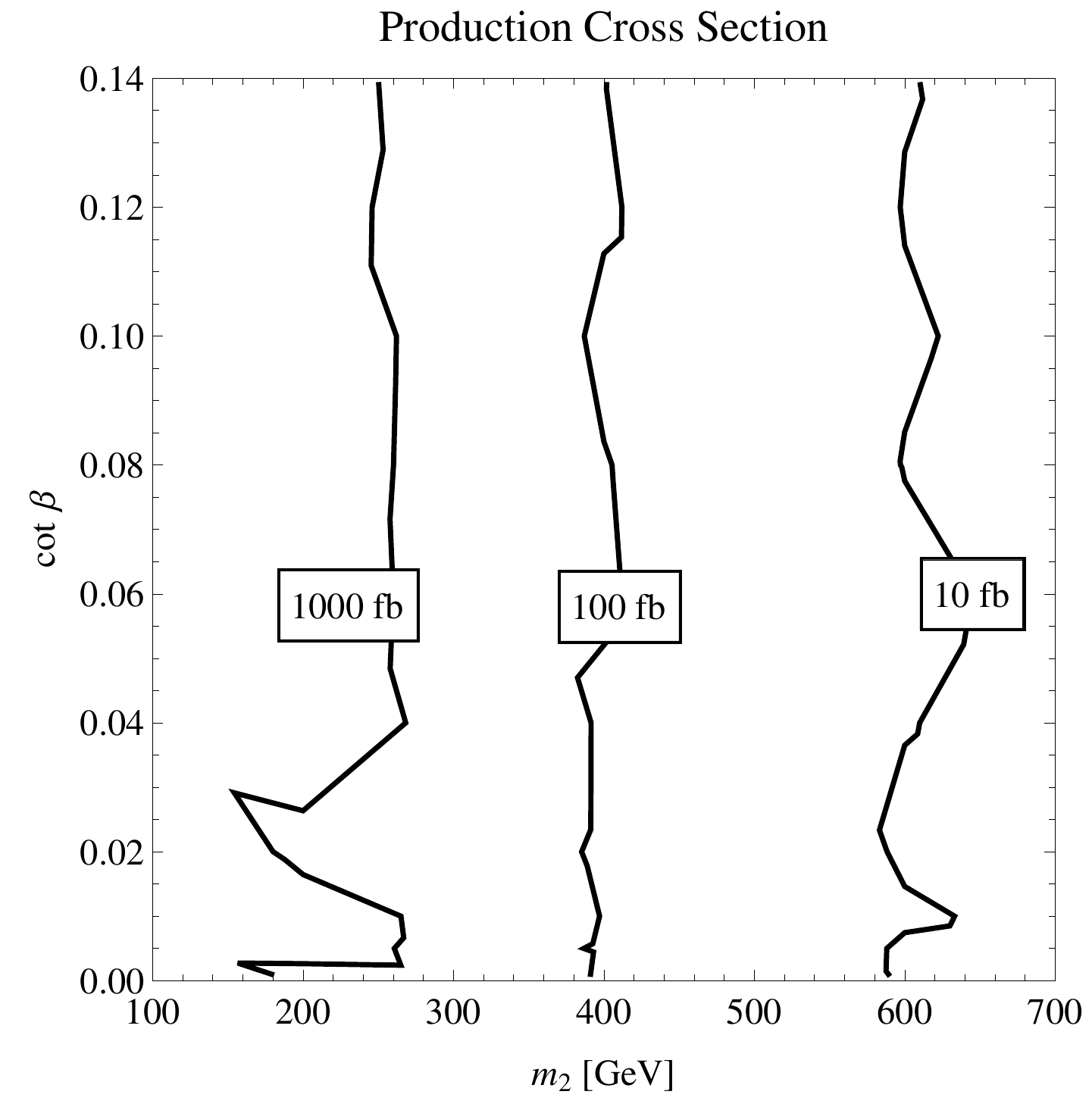}
        \end{subfigure}
         \caption{The total cross section in fb of the production of all possible pairs of septet particles.}
        \label{fig_CrossSectionandMostLim}
\end{figure}

The production of the multiply charged scalars depends only on their quantum
numbers, whereas production of the singly charged and neutral components depends
on the mixing between the septet and doublet. The $W$ boson couples most
strongly to the middle of the representation, while the $Z$ couples most
strongly to the top and bottom of the representation.  The coupling to the
photon is strongest for the particles with the highest charge. With these
observations and the mass hierarchy from Eq. \eqref{mass_hierarchy}, the order of
the cross sections tends to be
\begin{multline}
\sigma(pp\rightarrow H_{1,2}^-~H_{1,2}^+) >  \sigma(pp\rightarrow \chi^{++} H_{1,2}^-) > \sigma(pp\rightarrow \chi^{+3}\chi^{--}) \\> \sigma(pp\rightarrow \chi^{+5} \chi^{5-}) > \sigma(pp\rightarrow \chi^{+4} \chi^{3-}) > \sigma(pp\rightarrow \chi^{+4} \chi^{4-}) > ...
\label{eqn_CrossSections}
\end{multline}
where the choice of (1, 2) depends on the mixing involved. Due to the mixing in
both the singly charged and the neutral components, \eqref{eqn_CrossSections} is
not completely general. The ordering of the cross sections of the neutral and
singly charged particles may slightly change among themselves as well compared
to the ordering with the multiply charged states. However, the ordering of the
multiply charged states with respect to other multiply charged states will remain constant. The large quantum numbers and sheer number of states in the septet leads to relatively large inclusive cross sections. For example, with $m_2 \sim 250 \text{ GeV}$ the cross section is on the order of a pb. In comparison the production of the much lighter $W^+W^-$ at $\sqrt{s}=8 \text{ TeV}$ is $\sim 35 \text{ pb}$. The total cross section in fb of all combinations of septet pairs is shown in Figure \ref{fig_CrossSectionandMostLim}.
 
The decay of a multiply charged state goes to the next lowest charged state and
either a $W$ or a singly charged Higgs. However, the septet states tend to be nearly degenerate so the $W$ or charged Higgs will be off-sell. The doubly charged scalar can decay to two $W$ bosons, or a $W$ and a singly charged Higgs. 
The dominant decay depends on the masses as well as $\alpha$ and $\beta$. Once
the decay chain reaches the neutral septet state, the decay will have to
continue through mixing with the neutral doublet. As $\cot\beta \rightarrow 0$ and $\alpha \rightarrow 0$ 
the $H_0$ couplings to fermions and gauge bosons go to zero as seen in \eqref{eqn_hVV} and \eqref{eqn_hff}, leading to a stable particle at the end of the decay chain. As mentioned before, \cite{Earl:2013jsa} shows that models such as these can be ruled out from a dark matter perspective. However, even at our smallest value of $\cot\beta=0.001$, the width of $H_0$ is large enough to decay instantaneously. We again ignore the dark matter constraint and only study the collider results.

Due to the relative strength of the production of multiply charged particles and their decays, the search strategy will be to look for many $W$s. If the $W$ decays leptonically, the lepton is easy to identify, and the neutrino leads to missing energy. The other option is to have hadronic $W$ decays, producing many jets.

For each of the 90 model points in the grid, we simulate proton-proton collisions at $\sqrt{s}=8$ TeV. We then employ the same search strategies as done by the CMS and ATLAS experiments. The searches used are: the CMS search for anomalous production of events with three or more leptons \cite{CMS:2013jfa}; the ATLAS search for strongly produced SUSY particles in final states with two same sign leptons and jets \cite{ATLAS:2013tma}; the ATLAS search for charginos and neutralinos in events with three leptons and missing transverse momentum \cite{ATLAS:2013rla}; and the ATLAS search for squarks and gluinos in final states with jets and missing transverse momentum and no leptons \cite{TheATLAScollaboration:2013fha}.

We use the experimental search data to calculate the $95\%$ CL limit on the number of allowed events in a particular bin, denoted by
$s_{i,95}$. To compute $s_{i,95}$, we use the SM expected number of events ($b_i$), the
error in the SM expected number ($\sigma_{b,i}$), and the number of events
observed by the experiment ($n_i$). These values are plugged into the following
equation, which is then numerically solved for $s_{i,95}$:
\begin{equation}
\frac{\int \delta b_i \text{Gaus}(\delta b_i,\frac{\sigma_{b,i}}{b_i})
\times \text{Pois}(n_i |b_i (1+\delta b_i )+s_{i,95})}
{\int \delta b_i \text{Gaus}(\delta b_i,\frac{\sigma_{b,i}}{b_i})
\times \text{Pois}(n_i |b_i (1+\delta b_i ))}=0.05 .
\label{eqn_si95}
\end{equation}
To compute the expected limit instead of the observed limit, $b_i$ is used in
place of $n_i$. Using the Gaussian distribution is an attempt to take into
account the uncertainties.  However, many of the ATLAS papers present their own
values for $s_{i,95}$, which include systematics and other uncertainties.  In
all such cases, our calculation of $s_{i,95}$ resulted in a larger value than
that given by ATLAS.  A larger value means that more signal events are required to
exclude a model point at the $95\%$ CL. We therefore use our numbers, calculated
with \eqref{eqn_si95}, in order to
remain conservative and to keep a consistent statistical method between the
searches that present values for $s_{i,95}$ and those that do not.

\begin{table}[t]
\hspace{-2cm}
\begin{tabular}{|ccccccc|}
\hline
Signal Region & Cuts	& Observed	&	SM Expected & Error & Expected $s_{i, 95}$ & Observed $s_{i,95}$\\
\hline
ATLAS-CONF& 0 b-jets & & & & & \\
2013-007 &$N_{jets} \ge 3$& & & & &\\
SR0b& $E_T^{\text{miss}} > 150 \text{ GeV}$& 5 &7.5 & 3.3 & 11.7 & 8.6 \\
Two same&$m_T > 100\text{ GeV} $& & & & &\\
sign leptons& $m_{eff} >400 \text{ GeV}$ && & & & \\
\hline
ATLAS-CONF& $m_{SFOS}<81.2\text{ GeV}$ & & & & & \\
2013-035& or $m_{SFOS}>101.2\text{ GeV}$ & & & & &\\
SRnoZc& $E_T^{\text{miss}} > 75\text{ GeV}$&5 & 4.4 & 1.8 & 8.6 & 9.5 \\
& $m_T > 100 \text{ GeV} $& & & & &\\
Three leptons & $p_T \text{ 3}^{\text{rd}} \ell > 30 \text{ GeV}$& & & & &\\
\hline
\end{tabular}
\caption{Experimental search bins which are the most constraining bin for the greatest number of model points. 
	These two bins were the most significant for $\sim 2/3$ of the model points tested. The transverse mass is defined as $m_T = \sqrt{2\cdot 
	E_T^{\text{miss}}\cdot p_T^{\ell}\cdot(1-\cos \Delta\phi_{\ell,E_T^{\text{miss}}})}$. The variable $m_{eff}$ is the scalar sum
	 of the transverse momentum of the two leptons. For the three lepton search, $m_{SFOS}$ is the 
	 invariant mass of the same flavor opposite sign
	lepton pair which is closest to the $Z$ mass. }

\label{tbl_ExperimentValues}
\end{table}%

The number of events in a given bin is a function of the luminosity, the cross section, and the acceptance
\begin{equation}
s_i = \mathcal{L} ~ \sigma_i ~ \epsilon_i.
\label{eqn_c95}
\end{equation}
The detector simulations from PGS give the acceptance for each bin ($\epsilon_i$) which can be used 
with the maximum number of events allowed, $s_{i,95}$, to find the cross section which is excluded at the $95\%$ CL. The total production cross section of each model point
is compared to the bin which constrains the cross section to the smallest value. 
The model point is excluded if the cross section is larger than this value.

When computing the {\it expected} limits on the cross section, around 2/3
of the model points are most limited by the three lepton search. In this search all same flavor opposite sign lepton pairs are
required to be away from the $Z$ mass and contain large amounts of missing energy. The transverse momentum of the non-paired
lepton is required to be above 30 GeV, and the transverse mass is required to be above that of the $W$ mass. The transverse mass is defined as 
$m_T = \sqrt{2\cdot E_T^{\text{miss}}\cdot p_T^{\ell}\cdot(1-\cos \Delta\phi_{\ell,E_T^{\text{miss}}})}$. 
Most of the other model points are most constrained by a search for two same sign leptons. The two same sign lepton search
requires more missing energy and at least three jets none of which are tagged as b-jets. The transverse mass of the leading lepton is again
required to be above the mass of the $W$. Finally, the scalar sum of the transverse momentum of both leptons must be greater than 400 GeV.
The cut information for these bins is shown in Table \ref{tbl_ExperimentValues}. The observed number of events passing the two lepton search
is less than expected, which leads to a lower allowed cross section for new physics. Thus, when computing the {\it observed} limits, the two same sign
lepton search is most limiting for most of the events, followed by the three lepton search.

\begin{figure}[h]
\centering
        \begin{subfigure}[b]{0.45\textwidth}
                \includegraphics[width=\textwidth]{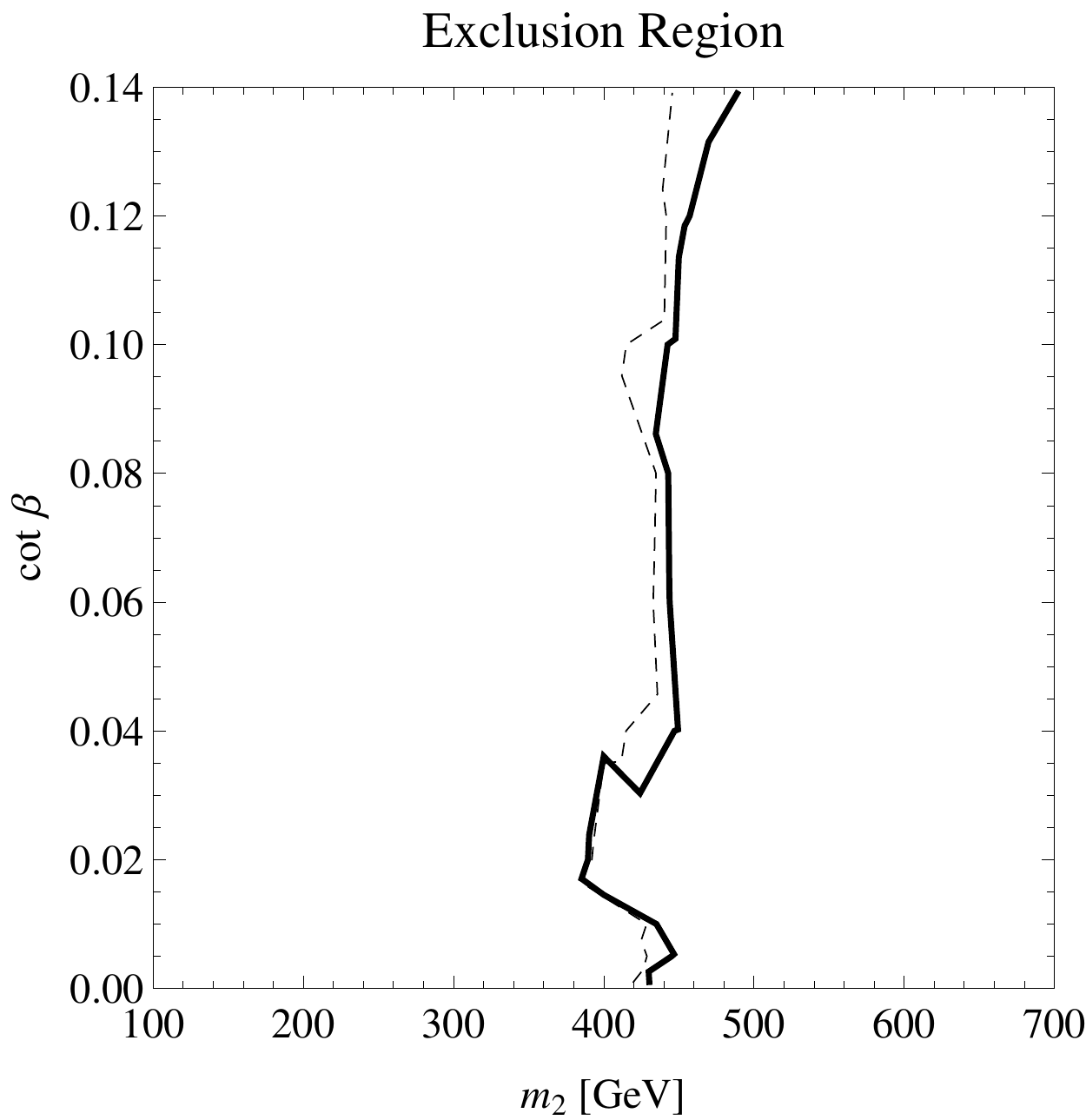}
        \end{subfigure}%

         \caption{A combination of ATLAS and CMS experimental searches for BSM physics 
		have been used to constrain the mass of the septet for fixed values of 
		the mixing angle ($\beta$) of the vevs of the septet and the doublet. 
		All model points follow the constraints from section \ref{section_Constraints}, specifically the tree level Higgs couplings
		do not change by more than $15\%$, the $h\rightarrow \gamma \gamma$ rate
		is consistent with observation, and the one loop contributions to the 
		oblique $S$ and $T$ parameters are within experimental limits. 
		The limits are calculated by considering allowed cross section for every bin and taking the smallest for each model point.
		The allowed cross section is obtained using equation \eqref{eqn_c95} where
		$s_{i,95}$ comes from equation \eqref{eqn_si95}. The dashed line represents the expected
		limit while the solid line is the observed limit. The model points to the left of the lines have a larger cross
		section than allowed by the experimental searches and are thus excluded.
		From this we have shown that LHC searches rule out a mass of the septet below about 400 GeV.}
        \label{fig_Exclusions}
\end{figure}

Figure \ref{fig_Exclusions} shows the limits coming from the most limiting bin at each model point. The dashed line is 
the expected limit, and follows the expected limit from the three lepton search. The solid line is the observed limit, which follows
that of the two same sign lepton search. The area to the left of the lines has a larger cross section than allowed
by the searches and is excluded. From this we rule out a mass parameter for the septet ($m_2$) 
below $\sim400$ GeV at the $95\%$ CL.

\section{Conclusion}
\label{sec_conclusion}
We have presented a phenomenological study of a general scalar septet extension
to the Standard Model, and shown that septet masses are excluded below about
400 GeV at the $95\%$ CL. This constraint came from using ATLAS and CMS searches
for BSM physics. The search for two same sign leptons gave the best observed limit followed 
closely by a three lepton search. The three lepton search gave the best expected limit.
The 400 GeV bound on $m_2$ is still low enough that $S$ and $T$ are not guaranteed to
be small, which happens closer to 1 TeV. This in turn forces small values of $\kappa_2$,
which is responsible for separating the mass of the different components of the septet.
Having a small value of $\kappa_2$ then allows the other doublet-septet coupling, $\kappa_1$
to determine the $h\rightarrow \gamma \gamma$ rate. For $m_2 \sim 400 $ GeV, a choice of $\kappa_1$
between -2 and 2 allows for a range of $h\rightarrow \gamma \gamma$ from 0 to above 10 times the standard model rate. Even if the mass
of the septet is pushed to 1 TeV, the septet model has enough charged particles coupling to the
Higgs that $\kappa_1$ has freedom to match a future more precise measurement of the diphoton coupling.

\paragraph{Acknowledgements} This research was supported in part by the 
Notre Dame Center for Research Computing through computing resources. 
We thank Joseph Bramante, Antonio Delgado, Adam Martin, and James Unwin for enlightening discussions. 
This work was partly supported by the National Science Foundation under grant PHY-1215979.


\begin{appendices}
	\section{Expanded Potential}
	\label{appendix_1}
As a reminder, the potential is given by
\begin{equation}
	\begin{aligned}
	V =& m_1^2 \Phi^2 + m_2^2 \chi^2 + \lambda (\Phi^{\dagger}\Phi)^2 
	-\frac{1}{\Lambda^3} \{(\chi^* \Phi^5 \Phi^*)+\text{H.C.}\}  \\
& + \sum_{A=1}^4 \lambda_A (\chi^{\dagger}\chi\chi^{\dagger}\chi)_A
+\sum_{B=1}^2 \kappa_B (\Phi^{\dagger}\Phi \chi^{\dagger}\chi)_B +\frac{1}{\Lambda^2} \sum_{C=1}^{3} \eta_C (\Phi^{\dagger~2}\Phi^2 \chi^{\dagger}\chi)_C.
	\end{aligned}
\end{equation}
For completeness, we here show the explicit tensor structure of the terms (summation
over repeated indices is assumed).
\begin{eqnarray}
(\chi^{\dagger}\chi \chi^{\dagger}\chi)_1 &= &\chi^{*~ijklmn}\chi_{ijklmn}\chi^{*~abcdef}\chi_{abcdef} \nn
(\chi^{\dagger}\chi \chi^{\dagger}\chi)_2 &=& \chi^{*~ijklmn}\chi_{ijklmf}\chi^{*~abcdef}\chi_{abcden}  \nn
(\chi^{\dagger}\chi \chi^{\dagger}\chi)_3 &=& \chi^{*~ijklmn}\chi_{ijklef}\chi^{*~abcdef}\chi_{abcdmn}  \nn
(\chi^{\dagger}\chi \chi^{\dagger}\chi)_4 &=& \chi^{*~ijklmn}\chi_{ijkdef}\chi^{*~abcdef}\chi_{abclmn}  \nn
(\Phi^{\dagger}\Phi \chi^{\dagger}\chi)_1 &=& \Phi^{*~i}\Phi_i \chi^{*~abcdef}\chi_{abcdef}  \\
(\Phi^{\dagger}\Phi \chi^{\dagger}\chi)_2 &=& \Phi^{*~i}\Phi_j \chi^{*~jabcde}\chi_{iabcde}  \nn 
(\chi^* \Phi^5 \Phi^*) &=& \chi^{*\,abcdef} \Phi_a \Phi_b \Phi_c \Phi_d \Phi_e \Phi^{*g} \epsilon_{fg} \nn
(\Phi^{\dagger~2}\Phi^2 \chi^{\dagger}\chi)_1 &=&  \Phi^{*~i}\Phi_i \Phi^{*~j}\Phi_j \chi^{*~abcdef}\chi_{abcdef} \nn
(\Phi^{\dagger~2}\Phi^2\chi^{\dagger}\chi)_2 &=&  \Phi^{*~i}\Phi_i \Phi^{*~j}\Phi_k \chi^{*~kbcdef}\chi_{jbcdef} \nn
(\Phi^{\dagger~2}\Phi^2 \chi^{\dagger}\chi)_3 &=&  \Phi^{*~i}\Phi_k
\Phi^{*~j}\Phi_l \chi^{*~lkcdef}\chi_{ijcdef} \nonumber
\end{eqnarray}
The field definitions in the tensor method are as follows.
\begin{eqnarray}
\begin{matrix} \Phi^1 = \Phi^+ \\ \Phi^2 = \Phi_0 \end{matrix} 
&& \begin{matrix} \chi^{111111} = \chi^{+5} \\ \chi^{211111} = 
	\chi^{+4}/\sqrt{6} \\ \chi^{221111} = \chi^{+3}/\sqrt{15} \\ 
	\chi^{222111} = \chi^{++} /2\sqrt{15} \\ \chi^{222211} = 
	\chi_1^+/\sqrt{15} \\ \chi^{222221} = 
	\chi^0/\sqrt{6} \\ \chi^{222222} = \chi_2^- \end{matrix}.
\end{eqnarray}

\section{$SU(2)$ Septet Generators}
\label{section_SU(2)Gen}
The septet is a (3,2) under $SU(2)_L\times U(1)_Y$. For convenience, the
generators for the 7-dimensional representation of $SU(2)$ are listed below,
along with $Q = T_3 + Y$.
\begin{equation}
	T_1 = \frac{1}{\sqrt{2}}
	\begin{pmatrix}
		0 & \sqrt{3} & 0 & 0 & 0 & 0 & 0 \\
		\sqrt{3} & 0 & \sqrt{5} & 0 & 0 & 0 & 0 \\
		0 & \sqrt{5} & 0 & \sqrt{6} & 0 & 0 & 0 \\
		0 & 0 & \sqrt{6} & 0 & \sqrt{6} & 0 & 0 \\
		0 & 0 & 0 & \sqrt{6} & 0 & \sqrt{5} & 0 \\
		0 & 0 & 0 & 0 & \sqrt{5} & 0 & \sqrt{3} \\
		0 & 0 & 0 & 0 & 0 & \sqrt{3} & 0 \\
	\end{pmatrix}
\end{equation}

\begin{equation}
	T_2 = \frac{i}{\sqrt{2}}
	\begin{pmatrix}
		0 & -\sqrt{3} & 0 & 0 & 0 & 0 & 0 \\
		\sqrt{3} & 0 & -\sqrt{5} & 0 & 0 & 0 & 0 \\
		0 & \sqrt{5} & 0 & -\sqrt{6} & 0 & 0 & 0 \\
		0 & 0 & \sqrt{6} & 0 & -\sqrt{6} & 0 & 0 \\
		0 & 0 & 0 & \sqrt{6} & 0 & -\sqrt{5} & 0 \\
		0 & 0 & 0 & 0 & \sqrt{5} & 0 & -\sqrt{3} \\
		0 & 0 & 0 & 0 & 0 & \sqrt{3} & 0 \\
	\end{pmatrix}
\end{equation}

\begin{equation}
	T_3 = 
	\begin{pmatrix}
		3 & 0 & 0 & 0 & 0 & 0 & 0 \\
		0 & 2 & 0 & 0 & 0 & 0 & 0 \\
		0 & 0 & 1 & 0 & 0 & 0 & 0 \\
		0 & 0 & 0 & 0 & 0 & 0 & 0 \\
		0 & 0 & 0 & 0 & -1 & 0 & 0 \\
		0 & 0 & 0 & 0 & 0 & -2 & 0 \\
		0 & 0 & 0 & 0 & 0 & 0 & -3 \\
	\end{pmatrix}
\end{equation}

\begin{equation}
	Q = 
	\begin{pmatrix}
		5 & 0 & 0 & 0 & 0 & 0 & 0 \\
		0 & 4 & 0 & 0 & 0 & 0 & 0 \\
		0 & 0 & 3 & 0 & 0 & 0 & 0 \\
		0 & 0 & 0 & 2 & 0 & 0 & 0 \\
		0 & 0 & 0 & 0 & 1 & 0 & 0 \\
		0 & 0 & 0 & 0 & 0 & 0 & 0 \\
		0 & 0 & 0 & 0 & 0 & 0 & -1 \\
	\end{pmatrix}
\end{equation}

\end{appendices}

\end{document}